\documentclass[aip,jcp,preprint,superscriptaddress]{revtex4-1}
\usepackage{amsmath}
\usepackage{latexsym}
\usepackage{mathrsfs}
\usepackage{graphicx}
\usepackage{dcolumn}
\usepackage{epsfig}
\usepackage{graphics}
\usepackage{color}
\usepackage{bm}
\usepackage{units,nicefrac}

\begin{document}

\title{Scaling law for crystal nucleation time in glasses}

\author{Anatolii V. Mokshin} \email{anatolii.mokshin@kpfu.ru}

\affiliation{Kazan Federal University, 420000 Kazan, Russia}
\affiliation{L.D. Landau Institute for Theoretical Physics,
Russian Academy of Sciences, 117940 Moscow, Russia}

\author{Bulat N. Galimzyanov} \email{bulatgnmail@gmail.com}

\affiliation{Kazan Federal University, 420000 Kazan, Russia}
\affiliation{L.D. Landau Institute for Theoretical Physics,
Russian Academy of Sciences, 117940 Moscow, Russia}

\date{\today}

\begin{abstract}
Due to high viscosity, glassy systems evolve slowly to the ordered
state. Results of molecular dynamics simulation reveal that the
structural ordering in glasses becomes observable over
``experimental'' (finite) time-scale for the range of phase
diagram with high values of pressure. We show that the structural
ordering in glasses at such conditions is initiated through the
nucleation mechanism, and the mechanism spreads to the states at
extremely deep levels of supercooling. We find that the scaled
values of the nucleation time, $\tau_1$ (average waiting time of
the first nucleus with the critical size), in glassy systems as a
function of the reduced temperature, $\widetilde{T}$, are
collapsed onto a single line reproducible by the power-law
dependence. This scaling is supported by the simulation results
for the model glassy systems for a wide range of temperatures as
well as by the experimental data for the stoichiometric glasses at
the temperatures near the glass transition.
\end{abstract}



\maketitle

\section{Introduction}

For a fluid supercooled isobarically  below the melting
temperature $T_m$, the ordered (say, crystalline) state is
thermodynamically favorable. At the moderate supercooling
$(T_m-T)/T_m$, the transition into an ordered state is started
through nucleation mechanism, that involves emergence of the
crystalline nuclei, which are able to grow; $T$ is temperature.
Hence, the behavior of the overall transition  should be
essentially determined by the rate characteristics: the waiting
time of the first crystalline critically-sized nucleus $\tau_1$,
the nucleation rate $J$ that is amount of the supercritical nuclei
formed per unit time per unit volume, and the growth rate $v_g$
that specifies growth law of the supercritical nuclei.

Some general features of the nucleation kinetics can be
comprehended within the classical nucleation
theory~\cite{Frenkel_1946,Turnbull_1949,Skripov_1974,Kelton_1991}
and its extensions~(see in
Refs.~\cite{Kashchiev_Nucleation_2000,Debenedetti_1996,Kalikmanov_review}).
According to the classical view, the driving force for the
nucleation grows upon the increase of supercooling. This means
that with the increase of supercooling an ordered state becomes
thermodynamically more favorable, while the waiting time for
nucleation $\tau_1$ and the nucleation time scale $1/J$ must be
shortened. On the other hand, with temperature lowering (below the
melting temperature $T_m$), the mobility of molecules (atoms)
decreases. As a result, any structural rearrangements, including
these responsible for nucleation, must be suppressed by the
growing viscosity. When a fluid is cooling down without
crystallization to temperatures corresponding to the viscosity
$\eta (T_g) \geq 10^{12}$ -- $10^{13}$~Pa$\cdot$s, it is
``freezing'' as disordered solid; where the temperature $T_g$ is
identified with the glass transition temperature. Although
crystallization of glasses proceeds over
time-scales~\cite{Saika_Voivod_PRL_2011,Mokshin/Barrat_PRE_2010},
which are commonly larger than experimentally acceptable, the
structural ordering in a glass can be accelerated by
out-of-equilibrium processes resulted from reheating or applied
shear
deformation~\cite{Mokshin/Galimzyanov/Barrat_PRE_2013,Heyes_JCP_2012,Mokshin/Barrat_PRE_2008,Kerrache_PRB_2011_1,Kerrache_PRB_2011_2}.
On the other hand, there are indications (see
Refs.~\cite{Durschang,Gutzow,Xing_JAP_2002,Yang_JPCM_2008,Niss_JCP_2008,Tanguy_2012})
that the time-scales of structural relaxation and of ordering in
glassy systems becomes shorter, when we move over
\emph{equilibrium} phase diagram to the range of more higher
pressures.

Moreover, debated issues in the field are related to the
temperature dependence of the transition rate characteristics at
deep levels of
supercooling~\cite{Zanotto_2002,Kalinina_1986,Muller_2000,Fokin/Zanotto_2003,Fokin/Yuritsyn/Zanotto_review,James_1989,Zanotto_1987,Deubener_2000}.
So, for example, empirical $T$-dependencies of the nucleation
lag-time and of the steady-state nucleation rate are discussed in
review~\cite{Fokin/Yuritsyn/Zanotto_review}, where results for
some stoichiometric glasses
($3$MgO$\cdot$Al$_2$O$_3\cdot3$SiO$_2$, Li$_2$O$\cdot2$SiO$_2$,
Na$_2$O$\cdot2$CaO$\cdot3$SiO$_2$, \textit{et al.}) are given. As
it is demonstrated in Ref.~\cite{Fokin/Yuritsyn/Zanotto_review}
within the available experimental data, the lag-time of nucleation
and the steady-state nucleation time-scale $1/J_s$ reach the
lowest values at a certain moderate levels of supercooling.
Moreover, both rate terms start to grow with the further increase
of supercooling and with the approaching the glass transition
temperature. Remarkably, the possible correlation discussed in
Ref.~\cite{Fokin/Yuritsyn/Zanotto_review} between some features in
the temperature dependencies of these rates (for example, the
maximum steady-state nucleation rate) and the reduced temperature
$T_g/T_m$  provides, in fact, indirect implications about
``unified laws'', which can be inherent in the nucleation
kinetics. In this work, we extend this view by focusing on the
crystal nucleation time $\tau_1$, identified here as the average
waiting time for the first critically-sized
nucleus~\cite{Kashchiev_Nucleation_2000,Shevkunov}.

The possibility of unified description using scaling relations has
been proposed and studied for the case of nucleation of liquid
droplets in \textit{the condensation process}. Here, an intriguing
feature emergent in the analysis of data for the vapor-to-liquid
nucleation is a supersaturation-temperature scaling of the
nucleation rate
data~\cite{Binder_scaling_1976,Hale_1,Hale_1_2,Hale_2}. Namely, as
shown by Hale~\cite{Hale_1,Hale_1_2,Hale_2}, data for the
nucleation rates plotted vs. $C_0 \ln S/[T_c/T -1 ]^{3/2}$ can
collapse onto a single line. Here, $S=p/p_{coex}$  is the
supersaturation, $p$ is the pressure of supersaturated vapor,
$p_{coex}$ is the pressure at the coexistence curve, $T_c$ is the
critical temperature, and $C_0$ is a normalization factor.  This
result is very interesting for the following reasons. First,
scaling relation allows one to compare the nucleation data of
various independent studies for a system, even though those
studies have not the identical pressure-temperature
(supersaturation-temperature) conditions. Moreover, if the scaling
is valid, then this is indication that there is a single reduced
variable instead of the pair, $T$ and $S$; and this variable is
sufficient for a unified description of the steady-state
vapor-to-liquid nucleation rate. Recently, Diemand \emph{et
al.}~\cite{Diemand_1,Diemand_2} suggested a new scaling relation
for the nucleation rate of homogeneous droplets from
supersaturated vapor phase, where other set of the parameters is
utilized. From the mentioned considerations, it is reasonable to
try to extend the ideas of scaling relations to the case of other
transition -- to the case of crystallization~\cite{Hale_3}. The
present study is mainly aimed at the consideration of this issue.
For this, the analysis of the crystal nucleation times from the
experiments and molecular dynamics simulations is carried out for
several systems at temperatures $T \leq T_g$.

The paper is organized as follows. In Sec.~\ref{Sec: Temp}, the
reduced temperature scale is introduced. Section~\ref{Sec:
sim_details} presents the simulation details and computational
methods. It includes the description of two model systems taken
for molecular dynamics simulations, the cluster analysis and the
statistical method utilized for the evaluation of the nucleation
characteristics within the simulation data. Discussion of the
results is given in Sec.~\ref{Sec: Discussion}. The main
conclusions are finally summarized in Sec.~\ref{Sec: Conclusion}.

\section{Reduced temperature scale \label{Sec: Temp}}

In evaluation of the unified temperature dependencies for
characteristics of the supercooled liquids, one encounters the
problem that the interested temperature range, $0\leq T \leq T_m$,
contains three control points  --  the zeroth temperature
$T=0$\;K; the glass transition temperature $T_g$ and the melting
temperature $T_m$, -- where $T_g$ and $T_m$ in the Kelvin scale
have not the same values for different systems. Therefore, there
is necessity to use a reduced temperature defined usually either
through $T_m$, or through $T_g$, depending on the
problem~\cite{Fokin/Zanotto_2003}.

For example, according to
Angell~\cite{Angell_1995,Angell_Leningrad_1989}, the inverse
reduced temperature $T_g/T$ is used to fulfill the
``strong-fragile'' classification of viscous (supercooled) liquids
by means of the plot, in which the viscosity in logarithmic scale,
$\log \eta$, is considered as a function of $T_g/T$. Since one has
$\log [\eta (T_g)] = 12$ -- $13$ for all the supercooled liquids
by definition, then the values of $\log \eta$ will be comparable
on the reduced temperature scale $0 < T_g/T \leq 1$ in
neighborhood of $T_g$. Further, the supercooling $(T_m-T)/T_m$ or
its conjugate quantity $T/T_m$ represent also the reduced
temperature scales (Ref.~\cite{Fokin/Yuritsyn/Zanotto_review}) and
are used to compare the characteristics of supercooled liquids for
temperature range $T \leq T_m$. Here, a reasonable consistency is
ensured as we approach the melting temperature $T_m$. Ambiguity of
the choice of reduce temperature scale is because the ratio
$T_g/T_m$ depends on the system (material) and can be different
even for the systems of same type. For example, the ratio of
$T_{g}/T_{m}$ for glasses Li$_2$O$\cdot2$SiO$_2$,
BaO$\cdot2$SiO$_{2}$ and 2Na$_{2}$O$\cdot$CaO$\cdot3$SiO$_{2}$,
which belong to the group of silicate glasses, does not have the
same value, and is equal to $0.56$, $0.568$, $0.512$,
respectively~\cite{Fokin/Zanotto_2003}. Moreover, the quantity
$T_g/T_m$ is dependent on cooling rate $dT/dt $ applied to prepare
glass at a desirable temperature and can have different values for
the different isobaric lines of a phase diagram. Therefore, the
absolute temperature $T$ as well as the reduced temperatures
$T/T_g$ and $T/T_m$ can not be considered as convenient
parameters, with respect to which evaluation of the unified
regularities could be examined.

\begin{figure}[!htbp]
\centering
\includegraphics[width=14cm,angle=0]{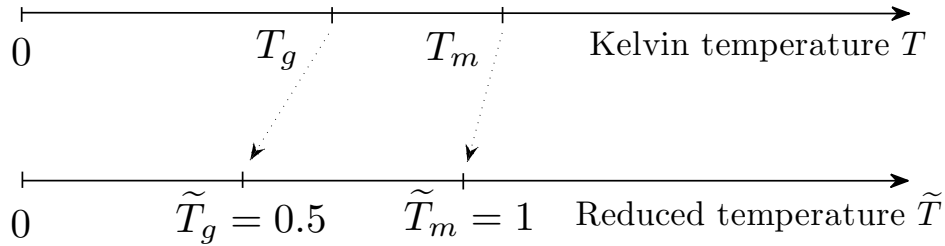}\vskip -1.5cm
\caption{ Demonstration of the transformation of the absolute
temperature scale $T$ into the reduced temperature scale
$\widetilde{T}$, which is system-independent and characterized by
fixed values of the melting temperature $\widetilde{T}_m=1$ and
the glass-transition temperature $\widetilde{T}_g=0.5$. Simple
realization of the transition can be done by means of
relation~(\ref{eq:system_omega_t}). \label{fig: temp}}
\end{figure}

To overcome this one needs to specify a temperature scale
$\tilde{T}$, in which the control points mentioned above -- the
zeroth temperature, the glass transition temperature and the
melting temperature -- are fixed and have same values for all
systems. We suggest a possible simple way to realize this. Let us
define the following correspondence between the values of
$\widetilde{T}$ for the three temperatures (the zeroth temperature
$T=0$\;K, the glass transition temperature $T_g$ and the melting
temperature $T_m$):
\begin{subequations} \label{eq: conditions}
\begin{equation}\label{eq:conditions2}
\widetilde{T} = 0 \hskip 0.5cm \mathrm{at} \hskip 0.5cm
T=0\;K,\end{equation}
\begin{equation}\label{eq:conditions2}
\widetilde{T}_{g}=0.5 \hskip 0.5cm \mathrm{at} \hskip 0.5cm T=T_g,
\end{equation}
\begin{equation}\label{eq:conditions2}
\widetilde{T}_{m}=1 \hskip 0.5cm \mathrm{at} \hskip 0.5cm
T=T_m.\end{equation}
\end{subequations}
The conditions~(\ref{eq: conditions}) are fulfilled with the
simple parabolic relation:
\begin{equation}
 \label{eq:system_omega_t}
\widetilde{T}=K_{1}\left ( \frac{T}{T_g}\right ) +K_{2} \left (
\frac{T}{T_g} \right )^{2}
\end{equation}
with
\begin{subequations}
\begin{equation}
K_1  + K_2 = 0.5,
\end{equation}
\vskip 0.1cm
\begin{equation} \label{eq: coef}
K_1 = \left ( \frac{0.5-
\displaystyle\frac{T_{g}^{2}}{T_m^2}}{1-\displaystyle\frac{T_{g}}{T_{m}}}\right
), \ \ \ K_2 = \left( \frac{
\displaystyle\frac{T_{g}}{T_m}-0.5}{\displaystyle\frac{T_{m}}{T_{g}}-1}\right
 ).
\end{equation}
\end{subequations}
With the known  $T_m$ and $T_g$ for a system,
relation~(\ref{eq:system_omega_t}) provides transform of the
absolute temperature scale $T$ into the reduced scale
$\widetilde{T}$, where all the temperature points coincide  for
all considered systems (see Fig.~\ref{fig: temp}). Moreover, when
the ratio $T_g/T_m$ approaches value $0.5$, the quadratic
contribution in Eq.~(\ref{eq:system_omega_t}) vanishes and
Eq.~(\ref{eq:system_omega_t}) is simplified to
\[
\widetilde{T} \simeq \frac{1}{2}\frac{T}{T_g}.
\]

It should be pointed out that the temperatures $T=0$\;K, $T_g$ and
$T_m$ in relation~(\ref{eq:system_omega_t}) correspond to the same
isobar. Relation~(\ref{eq:system_omega_t}) transforms  the ($p,T$)
phase diagram within the range $0 \leq T \leq T_m$ to the
($p,\widetilde{T}$) phase diagram unified for all systems, where
the glass transition line and the melting line are parallel to the
ordinate, $p$-axis, and intersect the abscissa at
$\widetilde{T}_g$=0.5 and $\widetilde{T}_m=1$, respectively.

\section{Simulation details and computational
methods \label{Sec: sim_details}}

In this work, we consider two model systems -- the Dzugutov (Dz)
system~\cite{Dzugutov_1992,Dzugutov_2002} and the binary
Lennard-Jones (bLJ)
mixture~\cite{Rowlinson_1969,Toxvaerd_Pedersen_2009}. Both systems
are known as the model glass-formers  suitable to study the
properties of glasses by means of molecular dynamics
simulations~\cite{Roth_PRB_2005,Kob_Andersen_1993,Hansen_McDonald_2006,Ryzhov}.
In this work, the glassy samples were generated by fast quench of
equilibrated fluids at the fixed pressure $p$.  The corresponding
pathways are shown on the phase diagrams in Fig.~\ref{fig:
phase_diagram}. Consideration of the ($p,T$)-points of the phase
diagrams with high values of the pressure $p$ allows us to deal
with such conditions at which the structural ordering in the
glassy  systems proceeds over time-scales available for
simulations even at temperatures below
$T_g$\;(Refs.~\cite{Mokshin/Barrat_PRE_2008,Mokshin/Barrat_PRE_2010}).
Thereby, the value of the pressure $p$ was chosen so that a
clearly-detected nucleation event was observable over the
simulation time scale. Hence,  the simulations with the generated
glassy samples were performed in the $NpT$-ensemble; $N$ is the
number of particles. The constant temperature and pressure
conditions are ensured by using the Nos\'{e}-Hoover thermostat and
barostat. For each ($p,T$)-point, more than fifty independent
samples were generated, the data of which were used in a
statistical treatment. For a single simulation run, $N=6\,912$
particles were enclosed in a cubic cell with periodic boundary
conditions. Note that the terms $\varepsilon$ and $\sigma$ define
the units of energy and length, respectively. Time, pressure, and
temperature units are measured in
$\tau_0=\sigma\sqrt{m/\varepsilon}$, $\varepsilon/\sigma^3$, and
$\varepsilon/k_B$, respectively.
\begin{figure}[!htbp]
\centering
\includegraphics[width=15cm,angle=0]{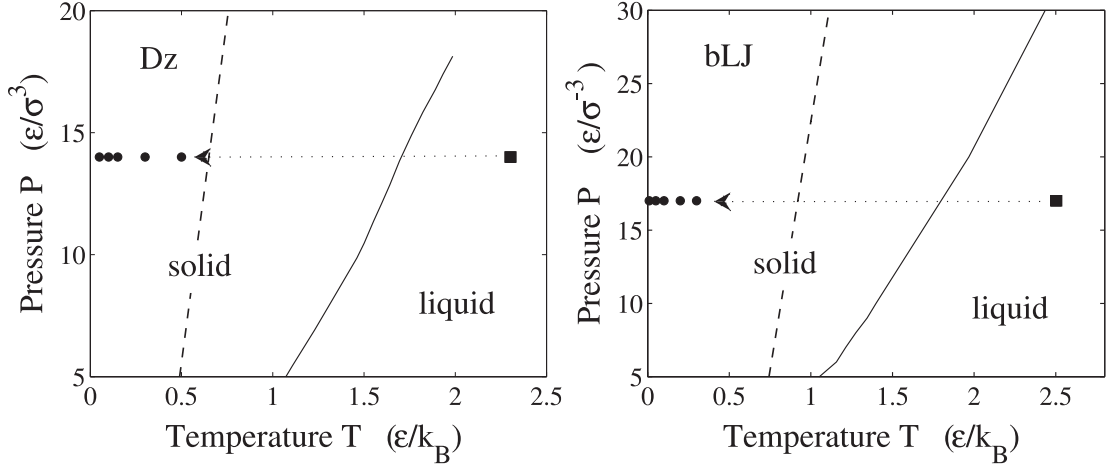}
\caption{ Pressure-temperature phase diagram for the Dz-system
(left panel) and for the bLJ-system (right panel). The full curves
denote the boundary between liquid and solid phases; the curve for
the Dzugutov system is reproduced from data of Fig.~4 in
Ref.~\cite{Roth_PRB_2005}. The dashed curves mark the boundary
between the supercooled liquid and the amorphous solid, when
liquid is cooled during isobaric simulations with the rate $d T/ d
t = 0.001$\;$\varepsilon/(k_B \tau_0)$. The full squares indicate
the equilibrium liquid states, which were used as starting points
to generate glassy samples. Pathways related with preparation of
the glassy samples are schematically shown by dotted arrows; and
the full circles denote the ($p,T$)-points, at which the
transition into ordered states was tracked. \label{fig:
phase_diagram}}
\end{figure}

\textit{The Dzugutov system.~--} In case of the Dzugutov system, all
particles are identical and interacting via a short-ranged pair
potential
\begin{eqnarray}
\label{eq:Dz_potential} \frac{U^{Dz}(r^*)}{\varepsilon} &=&
A({r^*}^{-m}-B)\exp\left(\frac{c}{r^*-a}\right)\Theta(a-r^*)\nonumber \\
&+& B\exp\left(\frac{d}{r^*-b}\right)\Theta(b-r^*),\nonumber
\\ & &
\hskip 0.3cm r^* = \frac{r_{ij}}{\sigma},
\end{eqnarray}
where $\Theta(\ldots)$ is the Heaviside step function, the values
of parameters $A=5.82$, $B=1.28$, $m=16$, $a=1.87$, $b=1.94$,
$c=1.1$ are chosen as suggested originally in
Ref.~\cite{Dzugutov_1992}. The simulations were performed for the
system along the isobaric line with the pressure
$p=14\,\varepsilon/\sigma^{3}$ at the temperatures $T=0.05$,
$0.1$, $0.15$, $0.3$ and $0.5\,\varepsilon/k_{B}$ below $T_g
\simeq 0.65\,\varepsilon/k_{B}$. For the isobar, the melting
temperature is $T_{m}\simeq1.51\,\varepsilon/k_{B}$, that yields
the temperature ratio $T_{g}/T_{m}\simeq0.43$.

\textit{The binary Lennard-Jones mixture.~--} The semi-empirical
(incomplete) Lorentz-Berthelot mixing
rules~\cite{Rowlinson_1969,Toxvaerd_Pedersen_2009},
\begin{eqnarray}
\sigma_{BB}&=&0.8\sigma_{AA},\nonumber \\
\sigma_{AB}&=&\frac{\sigma_{AA}+\sigma_{BB}}{2},\nonumber \\
\varepsilon_{BB}&=&0.5\varepsilon_{AA},\nonumber \\
\varepsilon_{AB}&=&\varepsilon_{AA}+\varepsilon_{BB},\nonumber
\end{eqnarray}
were utilized at the simulations of the binary Lennard-Jones
system $A_{80}B_{20}$ with the potential
\begin{equation}
\label{eq:BLJ_potential}
\frac{U_{\alpha\beta}^{bLJ}(r_{ij})}{\varepsilon_{\alpha\beta}}=4\left[\left(\frac{\sigma_{\alpha\beta}}{r_{ij}}\right)^{12}-
\left(\frac{\sigma_{\alpha\beta}}{r_{ij}}\right)^{6}\right],
\end{equation}
where $\alpha$, $\beta \in \{A,\, B\}$, the labels $A$ and $B$
denote the type of particles, $r_{ij}$ is the distance between the
centers of particles $i$ and $j$. Note that we take
$\varepsilon=\varepsilon_{AA}$, $\sigma=\sigma_{AA}$, and the mass
of a particle is $m=m_{A}=m_{B}=1$. For the bLJ system, we
consider the isobar with the  pressure
$p=17\,\varepsilon/\sigma^{3}$ at the temperatures $T=0.01$,
$0.05$, $0.1$, $0.2$ and $0.3\,\varepsilon/k_{B}$, that are lower
than the transition temperature
$T_{g}\simeq0.92\,\varepsilon/k_{B}$. The isobar contains the
melting point with $T_{m}\simeq1.65\,\varepsilon/k_{B}$.
Therefore, the temperature ratio is estimated as
$T_{g}/T_{m}\simeq 0.56$.

\textit{Cluster analysis.~--} The local domains of a crystalline
symmetry are examined by means of the cluster
analysis~\cite{Auer_Frenkel_2004,Mokshin_Barrat_2009}, introduced
originally by Wolde-Frenkel~\cite{Wolde-Frenkel}. The
consideration of the local environment around each particle is
performed by means of   $13$-dimensional complex vector with the
components~\cite{Steinhardt_Nelson_1983}
\begin{equation}
q_{6m}(i) = \frac{1}{N_b(i)} \sum_{j=1}^{N_b(i)}
Y_{6m}(\theta_{ij}, \varphi_{ij}).
\end{equation}
Here, $Y_{6m}(\theta_{ij}, \varphi_{ij})$ are spherical harmonics,
$N_b(i)$ is the number of neighbors for $i$ particle,
$\theta_{ij}$ and $\varphi_{ij}$ are polar and azimuthal angles,
which characterize the radius-vector $\vec{r}_{ij}$. Then, the
local order for each $i$ particle can be numerically evaluated by
means of the parameter~\cite{Steinhardt_Nelson_1983}
\begin{equation}
q_6(i) = \left (  \frac{4\pi}{13} \sum_{m=-6}^{6} |q_{6m}(i)|^2
\right )^{1/2},
\end{equation}
whereas degree of the orientational order can be estimated by
means of the global orientational order parameter $Q_6$ defined as
an average of $q_{6}(i)$ over all $N$
particles~\cite{Steinhardt_Nelson_1983}:
\begin{equation}
Q_6 = \frac{1}{N} \sum_{i=1}^{N} q_{6}(i).
\end{equation}
For a fully disordered system the parameter $q_6(i)$ is close to
zero, while it grows with increasing structural ordering. For
perfect fcc, bcc and hcp systems one has the largest possible
values for the parameters~\cite{Steinhardt_Nelson_1983}:
\[
q_6(i) = Q_6 \simeq 0.5745 \ \ \ \ \ \mathrm{(fcc)},
\]
\[
q_6(i) = Q_6 \simeq 0.5106 \ \ \ \ \ \mathrm{(bcc)},
\]
\[
q_6(i) = Q_6 \simeq 0.4848 \ \ \ \ \ \mathrm{(hcp)}.
\]

First, we define ``neighbors'' as all particles located within the
first coordination, the radius of which is associated with
position of the first minimum in the pair distribution
function~\cite{Mokshin/Barrat_PRE_2008}. Further, according to the
Wolde-Frenkel scheme~\cite{Wolde-Frenkel} we specify the pair of
neighboring particles ($i$ and $j$) as connected by a crystal-like
bond if the following condition is fulfilled:
\begin{equation}
\label{eq:cluster_analysis} 0.5 <
\left|\sum_{m=-6}^{6}\tilde{q}_{6m}(i)\tilde{q}_{6m}^{*}(j)
\right| \leq 1,
\end{equation}
where
\begin{equation}
\tilde{q}_{6m}(i)= \frac{{q}_{6m}(i)}{\left [
\displaystyle{\sum_{m=-6}^{6}} |{q}_{6m}(i)|^2 \right ]^{1/2}}.
\end{equation}
Condition~(\ref{eq:cluster_analysis}) allows one to distinguish
the particles correlated into an ordered
structure~\cite{Wolde-Frenkel}. Finally, particle $i$ is
identified as included into a crystalline structure if it has four
and more crystal-like bonds. The last condition is applied  to
exclude from consideration the structures with a negligible number
of bonds per particle, which occurs even in equilibrium liquid
phase~\cite{Mokshin/Barrat_PRE_2010}. By means of this routine,
the particles involved into the crystalline domains are detected.
\begin{figure}[!htbp]
\centering
\includegraphics[width=10.0cm,angle=0]{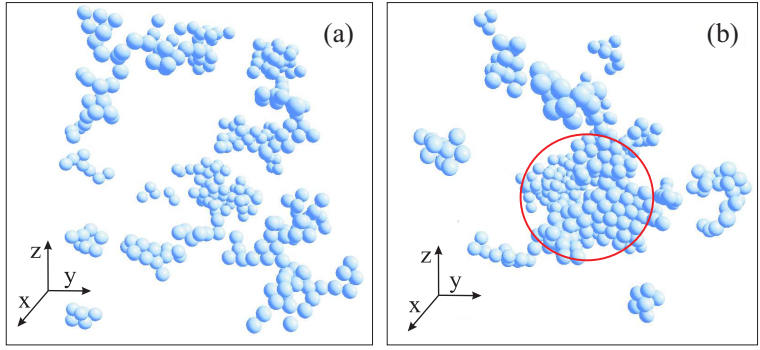}
\caption{(Color online) Snapshots of the Dz-system at
$T=0.5\,\varepsilon/k_B$ and at different times, for which the
particles recognized as belonging to the crystalline phase are
shown only. (a) System at the transient nucleation period;
$t=100\,\tau_0$. There are no nuclei capable to grow, and their
sizes are smaller than the critical size $n_c$. (b) System at the
time $t=250\,\tau_0$, when the first critically-sized nucleus
emerges; $n_c \simeq 105$~partilces. The critically-sized nucleus
is marked by red circle. \label{fig: snap}}
\end{figure}

Figure~\ref{fig: snap} demonstrates, as an example, the
crystalline clusters emerging in the glassy Dz-system at
$T=0.5\,\varepsilon/k_B$ over the transient nucleation regime,
where no nuclei capable to grow are detected [Fig.~~\ref{fig:
snap}(a)], and at the time $t=250\,\tau_0$, when the first nucleus
of the critical size appears [Fig.~~\ref{fig: snap}(b)].
\begin{figure}[!htbp]
\centering
\includegraphics[width=14cm,angle=0]{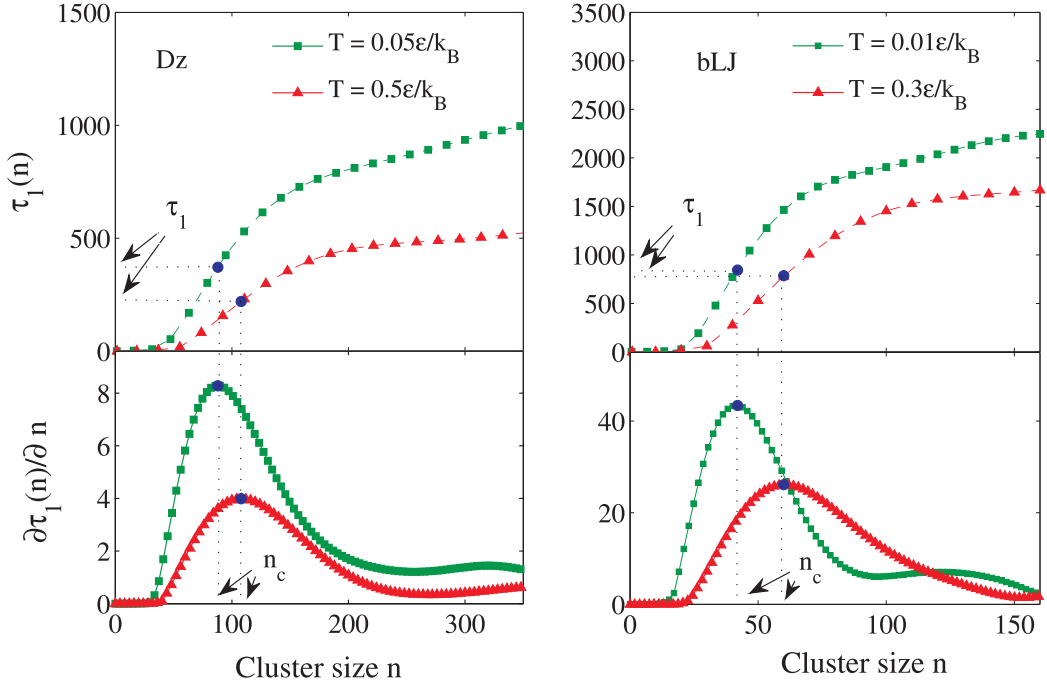}
\caption{(Color online) Mean-first-passage-time distributions
$\tau_1(n)$ and its first derivatives $\partial \tau_1(n)/\partial
n$ defined from simulation data for the Dz-system and the
bLJ-system at different temperatures. \label{fig: MFPT}}
\end{figure}

\vskip 0.5 cm \textit{Statistical treatment of the cluster
analysis results.~--} The growth trajectories of the crystalline
nuclei, $n_{\alpha i}(t)$, extracted from the different simulation
runs are treated within the mean-first-passage-time
method~\cite{Mokshin/Galimzyanov_JCP_2014,Mokshin_Galimzyanov_JPCB_2012}.
Here, $n$ defines number of the particles involved in the nucleus
at the time $t$, the mark $\alpha$ denotes the index of simulation
run, whereas the order number of the nucleation event $i$
indicates that the $i$th nucleus of the size $n$ appears at the
time $t$ during the $\alpha$th simulation run. On the basis of the
extracted trajectories $n_{\alpha i}(t)$, the
mean-first-passage-time distributions $\tau_i(n)$ are evaluated
for  each $i$th-order nucleus (for details, see
Ref.~\cite{Mokshin_Galimzyanov_JPCB_2012}). Further, the critical
size $n_c$ and the average waiting time for the $i$th-order
nucleus, $\tau_i$, $i=1,\, 2,\,\ldots$, are defined from the
analysis of the distributions $\tau_i(n)$ and of the first
derivatives $\partial \tau_i(n)/ \partial n$, according to the
scheme suggested in Ref.~\cite{Mokshin/Galimzyanov_JCP_2014}. In
this work, we focus on the characteristics for the largest nucleus
--  i.e. on its critical size $n_c$ and  average waiting time
$\tau_1$.

As an example, we show in Fig.~\ref{fig: MFPT} the
mean-first-passage-time distribution $\tau_1(n)$ and its first
derivative $\partial \tau_1(n)/\partial n$ computed for both the
systems.  As can be seen, the distributions $\tau_1(n)$ are
characterized by three regimes. The first regime, for which small
values of $n$ correspond to $\tau_1(n)$ with zero value, is
associated with pre-nucleation. Here, the nuclei with different
sizes (albeit, small sizes) appear with equal probability. The
second regime, in which the distribution $\tau_1(n)$ has the
pronounced non-zero slope, contains information about a nucleation
event. Namely, detected from the first derivative $\partial
\tau_1(n)/\partial n$  location of an inflection point in the
distribution $\tau_1(n)$ for the regime defines the critical size
$n_c$, whereas $\tau_1(n_c) \equiv \tau_1$ is directly associated
with the average waiting time of the first critically-sized
nucleus~\cite{Mokshin/Galimzyanov_JCP_2014}. Finally, the third
regime, where the slope of $\tau_1(n)$ decreases, corresponds to
growth of the nucleus. Note that such shape of the
mean-first-passage-time distribution is typical for an activated
process. The absence of the pronounced plateau in $\tau_1(n)$ for
the third regime indicates that the nuclei growth proceeds over a
time-scale comparable the nucleation time $\tau_1$
(Ref.~\cite{Mokshin_Galimzyanov_JPCB_2012}).
\begin{figure}[!htbp]
\centering
\includegraphics[width=15.5cm,angle=0]{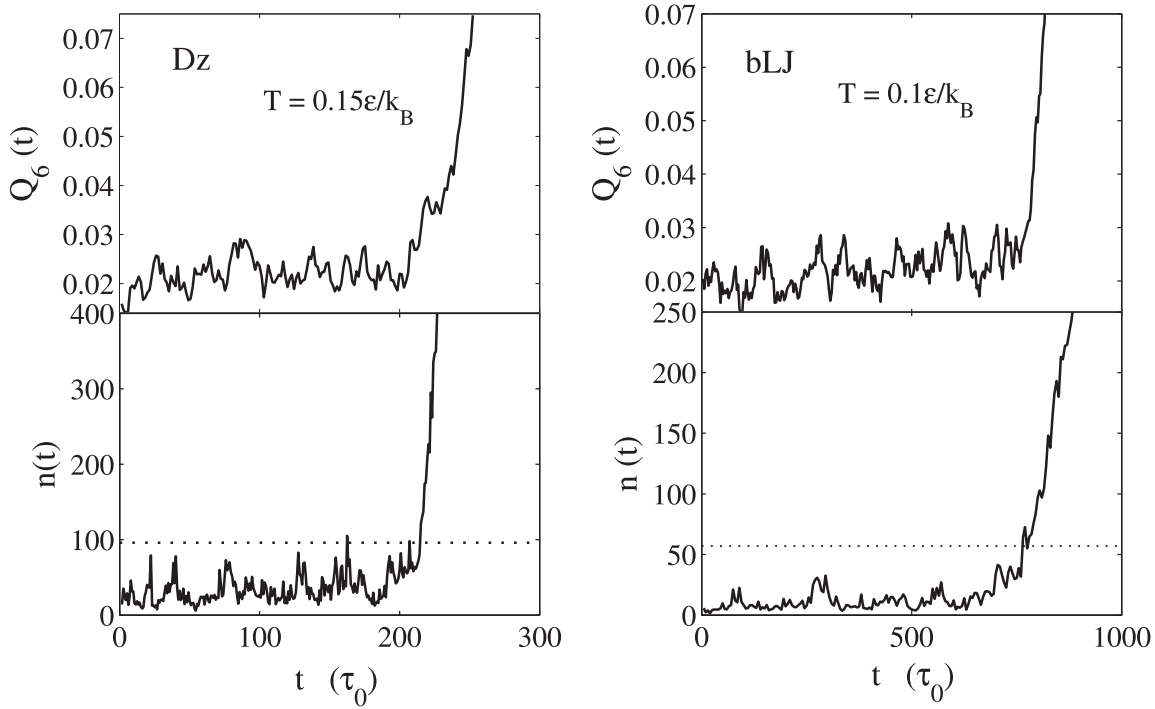}
\caption{ Trajectories of the global orientational order parameter
$Q_6(t)$ and of the  largest crystalline nucleus size $n(t)$
defined from a single simulation run for the Dz-system (left
panel) and the bLJ-system (right panel). The dotted horizonal
lines on the plots for $n(t)$ correspond to the critical sizes
$n_c$ defined from the statical analysis within the
mean-first-passage-time method. \label{fig: Q6}}
\end{figure}

\section{Discussion of results \label{Sec: Discussion}}

We start from evaluation of some properties of the nascent ordered
structures, that can help to elucidate the mechanism of the
ordering. Figure~\ref{fig: Q6} shows the time-dependent order
parameters -- the global orientational order parameter, $Q_6(t)$,
and the size of the largest cluster, $n(t)$, -- evaluated on the
basis of the simulation data. In initial stages, the parameters
$Q_6(t)$ and $n(t)$ fluctuate around their starting values. After
an incubation time, both the parameters start to growth rapidly.
Such evolution of the order parameters indicates on activated
character of the transition~\cite{Hanggi_RMP_1990}. The nucleation
event is clear detectable on a particular trajectory $n(t)$, where
it is associated with the start of sharp grow of $n(t)$. While
rough estimates for the nucleation time-scale $\tau_1$ and for the
critical size $n_c$ can be done even from the particular
trajectories $n(t)$ [see Fig.~\ref{fig: Q6}], the averaged values
for both the quantities can be computed directly by means of the
statistical method presented in Sec.~\ref{Sec: sim_details}.

Further, cluster analysis reveals that the nuclei of the critical
size are localized.  As contrasted to Ref.~\cite{Trudu/Parinello},
no ramified structures were detected even at very deep
supercooling. For quantitative characterization, the asphericity
parameter $S_0$ was computed according to
\begin{equation}
S_0 = \left \langle \frac{
(I_{xx}-I_{yy})^2+(I_{xx}-I_{zz})^2+(I_{yy}-I_{zz})^2 }{2
(I_{xx}+I_{yy}+I_{zz})^2} \right \rangle,
\end{equation}
where
\begin{equation}
I_{\alpha \beta} = \sum_{i=1}^{n_c} m (r_i^2 \delta_{\alpha \beta}
- r_{i\alpha}r_{i\beta})
\end{equation}
defines the components of the moment of inertia tensor associated
with a critically-sized nucleus; the brackets $\langle \ldots
\rangle$ mean the statistical average over results of the
different simulation runs. The parameter $S_0$ approximates the
unity, $S_0 \rightarrow 1$, for an elongated and ramified cluster,
and one has $S_0 \rightarrow 0$ for a cluster, the envelope of
which is of spherical shape. For both the systems (Dz and bLJ), we
find that the asphericity parameter is $S_0 \simeq 10^{-3}$, and
the size of the critical nucleus remains finite. This is evidence
that the transition into an ordered phase is initiated rather
through nucleation mechanism, and that is in agreement with
findings of Refs.~\cite{Saika_Voivod_JCP_2007,Bartell_JCP_2007}.
For the Dz-system, our estimations reveal that the critical size
changes from $n_c=108 \pm 5$  to $88 \pm 6$ particles with the
temperature decrease (increase of supercooling) from
$T=0.5\,\varepsilon/k_B$ to $0.05\,\varepsilon/k_B$. For the
bLJ-system,  we find that the critical size decreases from
$n_c=59\pm 4$ to $42 \pm 3$ particles with the temperature
decrease within the range $0.3\,\varepsilon/k_B \geq T \geq
0.01\,\varepsilon/k_B$.

\begin{figure}[!htbp]
\centering
\includegraphics[width=8.4cm,angle=0]{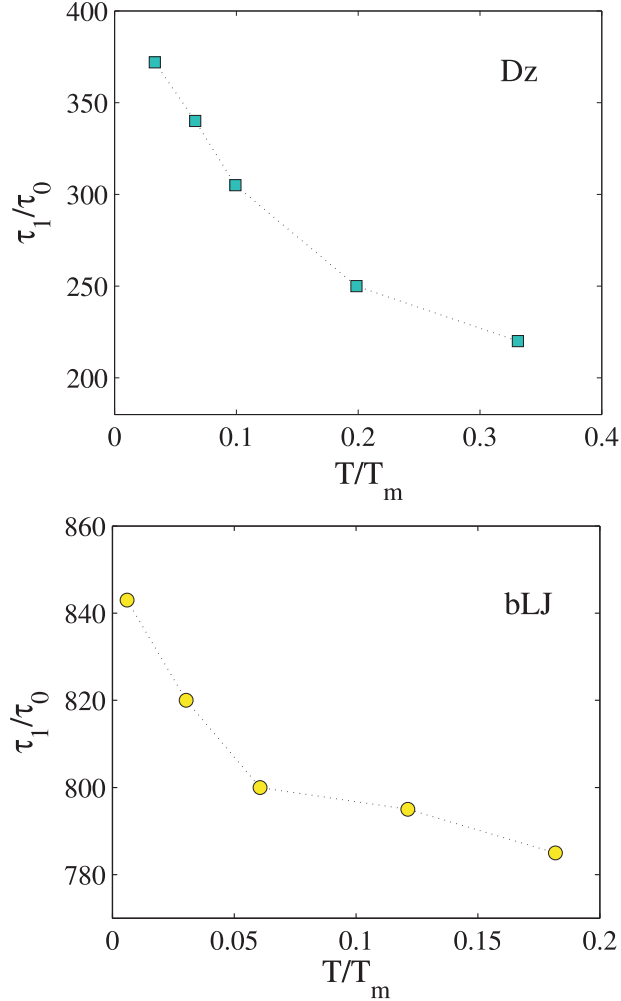}
\caption{(Color online) Average waiting time of the first
critically-sized nucleus $\tau_1$ (in units of $\tau_0$) versus
reduced temperature for the Dz-system ($T_g/T_m = 0.43$) and for
the bLJ-system ($T_g/T_m = 0.56$). The spanned thermodynamic
ranges correspond to deep levels of supercooling with the
temperatures below $T_g$. \label{fig: tau_1_raw}}
\end{figure}

Figure~\ref{fig: tau_1_raw} shows the values of the average
waiting time of the first nucleus of the critical size, $\tau_1$,
estimated from simulation data for the Dz-system and the
bLJ-system at the different temperatures. We note that the deep
levels of supercooling are considered for both the systems
corresponding to the temperatures much below $T_g$. The particle
mobility diminishing with supercooling results in the growth of
$\tau_1$ with the temperature decrease. The finite values of
$\tau_1$ comparable with the duration of numerical experiment may
seem surprising for a glassy system. Actually, the microscopic
kinetics of a glass changes with moving over phase diagram for the
range of high pressures~\cite{Tanguy_2012}. Namely, at high
pressures the structural relaxation as well as the transition of
glassy system into a state with the lower free energy proceeds
over shorter time
scales~\cite{Saika_Voivod_PRL_2011,Khusnutdinoff/Mokshin_PhA_2012,Khusnutdinoff/Mokshin_JNCS_2011}.
Therefore, the reduction of the values of $\tau_1$ is admissible
for the range of phase diagrams.

\begin{table}[ht]
\caption{The melting temperature $T_m$, the ratio $T_g/T_m$, the
waiting time for the first critically-sized nucleus $\tau_{1}^{g}$
at the transition temperature $T_g$,  the exponent $\gamma$
estimated from Eq.~(\ref{eq: reference_tau}), the parameters
$K_{1}$ and $K_{2}$ evaluated by Eq.~(\ref{eq: coef}) for several
systems.}
\begin{ruledtabular}
\begin{center}
\begin{tabular}{ccccccc}
System & $T_m$ & $T_g/T_m$ & $\tau_{1}^{g}$ & $\gamma$  & $K_{1}$ & $K_{2}$  \\
\hline
Dz (at $p=14\varepsilon/\sigma^3$)                       & $1.51\;\varepsilon/k_B$    & $0.43$ & $211\,\tau_0$ & $0.27$  & $0.553$ & $-0.053$ \\
bLJ (at $p=17\varepsilon/\sigma^3$)                      & $1.65\;\varepsilon/k_B$    & $0.56$ & $760\,\tau_0$ & $0.025$ & $0.427$ & $0.073$   \\
Li$_{2}$O$\cdot2$SiO$_{2}$     & $1286$\;K & $0.56$~\footnote{Experimental data of Ref.~\cite{Fokin/Zanotto_2003}.}  & $1869$\,sec~\footnote{From experimental data of Ref.~\cite{Fokin/Zanotto_2003}.} & $70$ & $0.424$ & $0.076$   \\
Na$_2$O$\cdot2$CaO$\cdot3$SiO$_2$ & $1549$\;K & $0.53$~\footnote{Experimental data of Ref.~\cite{Fokin/Zanotto_JNCS_2000}.}  & $5150$\,sec~\footnote{From experimental data of Ref.~\cite{Fokin/Zanotto_JNCS_2000}.} & $50$  & $0.466$ & $0.034$  \\
K$_2$O$\cdot$TiO$_2\cdot3$GeO$_2$ & $1308$\;K & $0.63$~\footnote{Experimental data of Ref.~\cite{Grujic_2009}.}  & $990$\,sec~\footnote{From experimental data of Ref.~\cite{Grujic_2009}.}  & $30$  & $0.281$ & $0.219$  \\
\end{tabular}
\label{Tab: parameters}
\end{center}
\end{ruledtabular}
\end{table}
Although the quantity $\tau_1$ for both the systems demonstrates
similar temperature dependence, it is difficult to say something
about quantitative  correspondence to the general nucleation
trends. Is such temperature dependence of the nucleation waiting
time, $\tau_1(T)$, is typical for the considered thermodynamic
range or not? One of the possible ways to clarify this is to bring
the extracted values of the nucleation waiting time $\tau_1(T)$
into a unified
scaled dependence.  
To construct scaling relation, we propose to use the reduced
temperature $\widetilde{T}$ defined by
relation~(\ref{eq:system_omega_t}), in which the values of the
glass transition temperature and the melting temperature are fixed
for all systems. Then, the simplest nonlinear
$\widetilde{T}$-dependence of $\tau_1$ can be chosen in the form:
\begin{equation} \label{eq: reference_tau}
\tau_1 = \tau_1^g \left( \frac{\widetilde{T}_g}{\widetilde{T}}
\right )^{\gamma},
\end{equation}
where $\tau_1^g$ is the average waiting time for the first
critically-sized nucleus at the state with the temperature $T_g$
(we remind that $\widetilde{T}_g = 0.5$). The dimensionless
parameter $\gamma > 0$ characterizes ability of the system at the
considered ($p,T$)-state to retain structural disorder. In
particular, the exponent $\gamma$ takes high values for the system
with good glass-forming properties, and must be characterized by
small values for the fast crystallizing systems. Since the
nucleation waiting time $\tau_1$ varies with pressure, then the
exponent $\gamma$ should be dependent on the pressure, at which a
supercooled liquid evolves. Namely, the exponent $\gamma$ is
decreasing function of the pressure $p$ for the systems, in which
the nucleation time scale decreases with pressure. The Dz and bLJ
systems correspond to the case.

\begin{figure}[!htbp]
\includegraphics[width=8.4cm,angle=0]{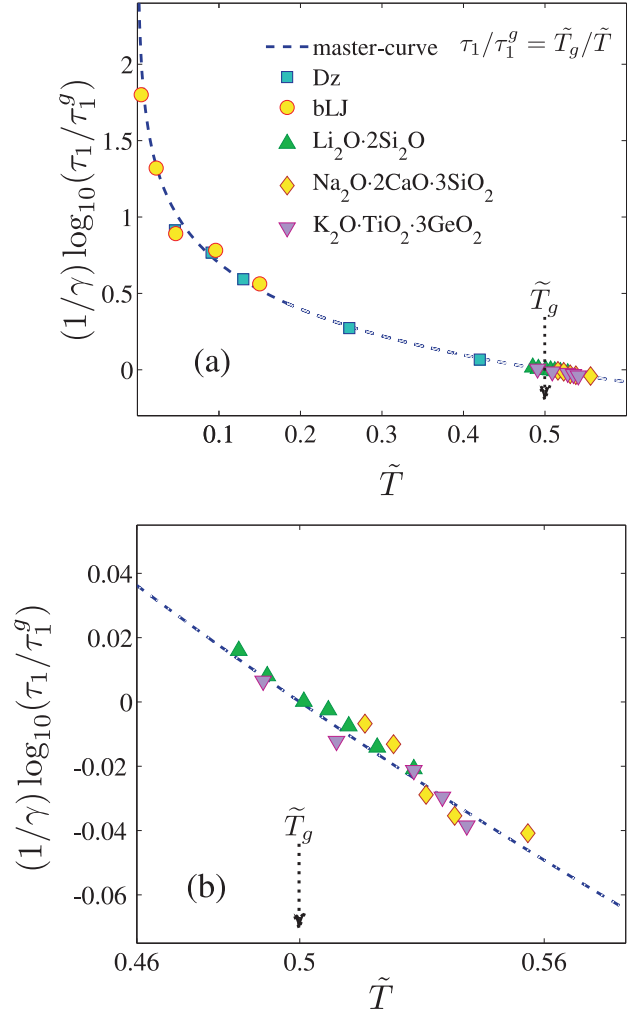}
\caption{(Color online)  (a) Scaled waiting time for the first
critically-sized nucleus $(1/\gamma)\log_{10}(\tau_1/\tau_1^g)$ is
plotted as a function of the reduced temperature $\widetilde{T}$.
Here, $\widetilde{T}_g = 0.5$ is the scaled transition temperature
(marked by arrow), $\tau_1^g$ is the waiting time at the
transition temperature ${T}_g$, and $\gamma$ is the fitting
parameter.  (b) The same but for the temperature range $0.46 \leq
\widetilde{T}\leq 0.58$. Values of the parameters $\tau_1^{g}$ and
$\gamma$ are given in Tab.~\ref{Tab: parameters}.
 Consistency of the data to the master-curve, which set the
temperature-dependence $\widetilde{T}_g/\widetilde{T}$, provides
support to the validity of scaling relation~(\ref{eq:
reference_tau}).\label{fig: tau1_scaled}}
\end{figure}

To verify validity of relation (\ref{eq: reference_tau}), we place
the rescaled data for the average waiting time $\tau_1$ for the Dz
and bLJ systems vs. the reduced temperature $\widetilde{T}$ on the
common Fig.~\ref{fig: tau1_scaled}. For clarity, the axis of
ordinates is presented on a logarithmic scale, where the fitting
parameter $\gamma$ corrects the slope in accordance with the
master-curve
\begin{equation} \label{eq: master_curve}
\left ( \frac{\tau_1}{\tau_1^g} \right ) =
\frac{\widetilde{T}_g}{\widetilde{T}},
\end{equation}
which appears from (\ref{eq: reference_tau}) at the exponent
$\gamma=1$. The reduced temperature $\widetilde{T}$ in
Eq.~(\ref{eq: reference_tau}) guarantees that the temperature
points spread over the abscissa in the same manner for all the
considered systems, whereas the dimensionless parameter $\gamma$
forces all the ordinate points to collapse onto the
master-curve~(\ref{eq: master_curve}). Since our simulation
results for the Dz and bLJ systems cover the temperature range $T
< T_g$ and  we did not estimate the nucleation time $\tau_1$ at
the transition temperature $T_g$, then the term $\tau_1^g$ was
taken as a fitting parameter. Namely, its values were found by
extrapolation of the data for $\tau_1$ to the temperature point
$\widetilde{T}_g=0.5$, where the function
$(1/\gamma)\log(\tau_1/\tau_1^g)$ must be equal to zero (see
Fig.~\ref{fig: tau1_scaled}). Numerical values of $\tau_1^g$ are
given in Tab.~\ref{Tab: parameters}. As can be seen from
Fig.~\ref{fig: tau1_scaled}, all the data obtained on the basis of
molecular dynamics simulations follow the unified master-curve.
Moreover, in contrast to the case of the Kelvin temperature scale,
values of the melting temperature $\widetilde{T}_m$ and of the
transition temperature $\widetilde{T_g}$ are not dependent on
pressure. Therefore, the results shown on Fig.~\ref{fig:
tau1_scaled} can be supplemented by the data for any supercooled
liquid at arbitrary value of the pressure $p$.

Moreover, it is attractive to extend the study and to verify the
scaling law~(\ref{eq: reference_tau}) with the experimental data.
While the direct experimental measurements of $\tau_1$ are
difficult~\cite{Shneidman}, we suggest the next routine for the
approximative estimation of $\tau_1$, which can be realized with
the experimentally measurable quantities -- the steady-state
nucleation rate $J_s$ and the induction time $\tau_{ind}$.
According to
Kashchiev~\cite{Kashchiev_1969,Kashchiev_Nucleation_2000}, the
number density of the supercritical nuclei in the system, $i_{V}$,
evolves with time as
\begin{eqnarray} \label{eq:nvt_expression1}
\frac{i_{V}(t)}{J_{st}\tau_{ind}} &=&
\frac{t}{\tau_{ind}}-1  \\
&-&
\frac{12}{\pi^{2}}\sum_{m=1}^{\infty}\frac{(-1)^{m}}{m^{2}}\exp\left(-\frac{m^{2}\pi^{2}t}{6\tau_{ind}}\right).\nonumber
\end{eqnarray}
For the time $t=\tau_1$ one has $i_V(\tau_1)= 1/V$,  and
Eq.~(\ref{eq:nvt_expression1}) takes the form:
\begin{eqnarray} \label{eq: simplification}
\frac{1}{J_{st}V} &=& \tau_1 - \tau_{ind}  \\ &+& \frac{12
\tau_{ind}}{\pi^{2}}\sum_{m=1}^{\infty}\frac{(-1)^{m}}{m^{2}}\exp\left(-\frac{m^{2}\pi^{2}\tau_1}{6\tau_{ind}}\right).\nonumber
\end{eqnarray}
Further, Eq.~(\ref{eq: simplification}) was numerically solved
with the experimental $J_s$ and $\tau_{ind}$ for
Li$_{2}$O$\cdot2$SiO$_{2}$ reported in
Ref.~\cite{Fokin/Yuritsyn/Zanotto_review}, for
Na$_2$O$\cdot2$CaO$\cdot3$SiO$_2$ presented in
Ref.~\cite{Fokin/Zanotto_JNCS_2000}, and for
K$_2$O$\cdot$TiO$_2\cdot3$GeO$_2$ given in
Ref.~\cite{Grujic_2009}. The extracted rescaled values of the
average waiting time $\tau_1$ are also presented in Fig.~\ref{fig:
tau1_scaled}. As can be seen from Fig.~\ref{fig: tau1_scaled},
``experimental'' data for $\tau_1$ provide the $T$-dependence,
which is in agreement with scaling relation~(\ref{eq:
reference_tau}) as well as with the simulation results for the
bLJ-system and the Dz-system.

Analysis of the reduced temperature scale $\widetilde{T}$ for the
systems reveals that the quadratic contribution in equation for
$\widetilde{T}$ [see Eq.~(\ref{eq:system_omega_t})] can be
insignificant as for the Dz-system and for
Na$_2$O$\cdot2$CaO$\cdot3$SiO$_2$, where the ratio $T_g/T_m$ is
equal to $0.43$ and $0.53$, respectively (see Tab.~\ref{Tab:
parameters}). With away from value $0.5$ for the ratio $T_g/T_m$,
weight of the quadratic contribution, $K_2$, increases. The values
of the parameters $K_1$ and $K_2$ are comparable for
K$_2$O$\cdot$TiO$_2\cdot3$GeO$_2$ characterized by the ratio
$T_g/T_m=0.63$. To our knowledge, the highest value of the ratio
$T_g/T_m$ appears for Na$_2$O$\cdot$Al$_2$O$_3\cdot6$SiO$_2$ and
is equal to $0.78$~(Ref.~\cite{Fokin/Yuritsyn/Zanotto_review}).

The values of the exponent $\gamma$ differ for the considered
systems by four orders of magnitude, and an order of magnitude
between the Dz-system and the bLJ-system (see Tab.~\ref{Tab:
parameters}). The large scatter in the values of $\gamma$ is due
to the change of the waiting nucleation time $\tau_1$ within the
temperature range differs essentially for the systems. One can
demonstrate this with the results for the Dz and bLJ systems shown
on Fig.~\ref{fig: tau_1_raw}. Within the temperature range $0.025
\leq T/T_m \leq 0.2$ the time scale $\tau_1$ is changed by the
factor $0.625$ for the Dz-system, whereas it changes by the factor
$0.96$ for the bLJ-system. For the systems with complicated
structural units (i.e. for silicate glasses) the change is much
more pronounced~\cite{Fokin/Yuritsyn/Zanotto_review}. The case
with the smaller change in the temperature dependence of $\tau_1$
will corresponds to the smaller values of the exponent $\gamma$ in
scaling relation~(\ref{eq: reference_tau}).

Although the scaling law~(\ref{eq: reference_tau}) is suggested
rather as an empirical result, its qualitative  justification can
also be done. At the temperatures comparable with and lower than
the glass transition temperature $T_g$, the local structural
rearrangements responsible for the nucleation are driven rather by
kinetic aspects associated with the viscosity than by
thermodynamic contribution. Therefore, it is reasonable for the
range of high supercooling to expect the existence of correlation
between  the waiting time for nucleation $\tau_1(T)$ and the
structural relaxation time $\tau_{\alpha}(T) \sim \eta(T)$, and,
thereby, between the time $\tau_1(T)$ and the viscosity $
\eta(T)$:
\begin{equation} \label{eq: prop}
\tau_1(T) \sim \eta(T).
 \end{equation}

Hence, the Vogel-Fulcher-Tammann equation provides the most
popular viscosity model (this equation is also known as the
Williams-Landel-Ferry model~\cite{Seeton,Allan_PNAS}):
\begin{equation} \label{eq: VFT}
\log_{10} {\eta}(T) = \log_{10} {\eta}_{\infty} + \frac{A}{T-T_c},
\end{equation}
where $T_c$ is the critical temperature of this model. Another
equation for viscosity similar to VFT-model is provided by the
mode-coupling theory~\cite{Angell_Leningrad_1989,Gotze_book_2009}:
\begin{equation} \label{eq: MCT}
\eta(T) = \frac{{\eta}_{\infty}}{(T-T_{MCT})^{\gamma_m}},
\end{equation}
where  $T_{MCT}$ is the (critical) mode-coupling temperature. The
parameters ${\eta}_{\infty}$, $A$ and $\gamma_m$ take positive
values and are obtained by fitting Eqs.~(\ref{eq: VFT}), (\ref{eq:
MCT}) to experimentally measured viscosity data~\cite{Allan_PNAS}.
Both the models predict a divergence of the viscosity $\eta(T)$
when $T \rightarrow T_c$ (and $T \rightarrow T_{MCT}$). Moreover,
both the models are able to reproduce $\eta(T)$ for the
supercooled liquid phase, i.e. $T>T_c$ (and $T > T_{MCT}$), and
are not applicable for the temperature range below $T_c$ (below
$T_{MCT}$) because of a divergence in the temperature
dependencies. On the other hand, for a high-viscosity regime
corresponding to the temperatures $T \leq T_g$, the experimentally
measured temperature-dependence of the viscosity $\eta(T)$ is
reproducible by the Arrhenius law (see, for example, Fig.~6 in
Ref.~\cite{Bottinga}), which is generalized by the Avramov-Milchev
equation~\cite{Avramov/Milchev_1988,Allan_PNAS}:
\begin{equation}
\log_{10}\eta(T) = \log_{10}\eta_{\infty} + \left (
\frac{\mathcal{A}}{T} \right )^{\alpha '}
\end{equation}
or
\begin{equation} \label{eq: viscosity_AM}
\log_{10}\left [ \frac{\eta(T)}{\eta_{\infty}} \right ]= \left (
\frac{T}{\mathcal{A}} \right )^{-\alpha '},
\end{equation}
where $A$ and $\alpha '$ are positive.

On the other hand, let us now reconsider scaling
relation~(\ref{eq: reference_tau}), which can be rewritten in the
form
\begin{equation} \label{eq: tau_1}
\frac{\tau_1}{\tau_1^g} = \left(2 \widetilde{T} \right
)^{-\gamma},
\end{equation}
since $\widetilde{T}_g=0.5$. After substitution of
Eq.~(\ref{eq:system_omega_t}) into relation~(\ref{eq: tau_1}) and
using the expansion
\begin{equation}
\ln(x+1) = \sum_{n=1}^{\infty} \frac{(-1)^{n-1}x^n}{n}, \hskip 1cm
-1 < x < 1,
\end{equation}
we obtain for the temperature range  $0 < T < T_m$ the following
equation:
\begin{eqnarray} \label{eq: expansion_tau}
\log_{10} \left [ \frac{\tau_1(T)}{\tau_1^g} \right ] &=&
\frac{1}{\ln{10}}\sum_{n=1}^{\infty} \frac{(-1)^{n-1}}{n} \\ & &
\times  \left ( \frac{T}{T_g}  \right )^{-\gamma n } \left [ \left
\{ 2K_1 \left (1- \frac{T}{T_g}\right ) +\frac{T}{T_g} \right
\}^{-\gamma} - \left ( \frac{T}{T_g} \right )^{\gamma} \right
]^{n},  \nonumber
\end{eqnarray}
where the parameter $K_1$ is defined by Eq.~(\ref{eq: coef}).
Assuming that proportionality in Eq.~(\ref{eq: prop}) holds, one
can compare r.h.s. of Eqs.~(\ref{eq: viscosity_AM}) and (\ref{eq:
expansion_tau}). A simple analysis reveals that Eq.~(\ref{eq:
expansion_tau}) is able to approximate the power-law dependence of
Eq.~(\ref{eq: viscosity_AM}) and generalizes the temperature
dependence for the viscosity given by the Avramov-Milchev
equation. Thereby, the scaling relation~(\ref{eq: reference_tau})
and the viscosity model with Eq.~(\ref{eq: viscosity_AM}) can be
considered as consistent.

Moreover, the fragility of a system can be estimated by means of
the index $m$ defined as~\cite{Angell_1995}
\begin{equation} \label{eq: fragility_Angell}
m = \left .     \frac{\partial \log_{10} (\eta) }{\partial (T_g/T)
} \right |_{T=T_g} .
\end{equation}
Then, from Eqs.~(\ref{eq: prop}),
 (\ref{eq: tau_1}) and (\ref{eq: fragility_Angell})  we obtain the following relation
\begin{equation} \label{eq: gamma_fragility}
m \sim \left .     \frac{\partial \log_{10} (\tau_1) }{\partial
(T_g/T) } \right |_{T=T_g} \sim 2\gamma (1-K_1),
\end{equation}
which after substitution of Eq.~(\ref{eq: coef}) can be rewritten
as
\begin{equation} \label{eq: frag_2}
m \propto 2 \gamma \left [  \frac{\displaystyle 0.5 -
\frac{T_g}{T_m}+\left ( \frac{T_g}{T_m} \right )^2 }{\displaystyle
1- \frac{T_g}{T_m} } \right ].
\end{equation}
Here the contribution in square brackets is positive for the range
$0 \leq (T_g/T_m) \leq 1$. Last two relations indicates that the
exponent $\gamma$ and the index $m$ are correlated terms, whereas
$\gamma$ can provide an estimate of fragility.

\section{Conclusion \label{Sec: Conclusion}}

The mechanism of the structural ordering in the supercooled melts
at extremely deep level of supercooling is one of the most debated
issues in the consideration of the crystallization
kinetics~\cite{van_Megen_Nature_1993,Cavagna_2003,Cavagna_2007}.
Let us mention some viewpoints in this regard. The mean-field
theories, starting from the gradient theory of Cahn-Hilliard,
provide indications that the structural ordering at a deep level
of metastability can proceed through the spinodal
decomposition~\cite{Cahn/Hilliard_JCP_1959}. Interestingly, Trudu
\textit{et al.} for the freezing bulk Lennard-Jones system found a
spatially diffuse and collective phenomenon of nucleation at deep
supercooling. Authors treated such features as indirect signatures
of a mean-field spinodal~\cite{Trudu/Parinello}. This was later
criticized by Bartell and Wu~\cite{Bartell_JCP_2007}. According to
experimental~\cite{Pan_2006} and other
simulation~\cite{Bagchi_2007,Saika_Voivod_JCP_2007} studies, the
size of the critical embryo remains finite with decrease of the
temperature of the supercooled liquid, in contrast to the
mean-field theory predictions for a spinodal. Moreover, results of
Ref.~\cite{Sanz_PRL_2011} reveal that crystallization in hard
sphere glasses proceeds due to ``a chaotic sequence of random
micronucleation events, correlated in space by emergent dynamic
heterogeneity'', and agree with findings of
Bartell-Wu~\cite{Bartell_JCP_2007}. In view of this, it remains
still desirable to examine  the mechanisms of the structural
ordering in glasses within the new experimental/simulation
results.

In the present work,  two model glassy systems with different
interparticle interaction  -- the single-component Dzugutov system
and the binary Lennard-Jones system -- are simulated with the aim
to study the structural ordering at deep supercooling. Remarkably,
the simulation study covers a wide temperature range: from the
temperatures comparable with $T_g$ to the temperatures
corresponding to very deep levels of supercooling $(T_m - T)/T_m
\simeq 0.97$.  By means of cluster analysis, we show that the
structural ordering even at deep supercooling proceeds through the
formation of the localized crystalline domains, where the size of
the critical embryo still remains finite. This supports the
nucleation scenario of crystallization in the glassy systems, and
is in agreement with the recent findings of Saika-Voivod
\textit{et al.}
~\cite{Saika_Voivod_PRL_2011,Saika_Voivod_JCP_2007} and Sanz
\textit{et al.}~\cite{Sanz_PRL_2011}.

The average nucleation time is the quantity of main interest in
the characterization of the initial stages in the nucleation
kinetics. Here, it is estimated on the basis of the molecular
dynamics simulation data (for the two model glassy systems) and
from the available experimental data for the several glasses
within the Kashchiev's approximative equation. Our results show
that, with the decrease of the temperature,  the nucleation time
$\tau_1$ increases but still remains finite. Further, we find that
the nucleation time $\tau_1$ plotted as a function of the proposed
reduced temperature  follows the power-law dependence, unified for
all the considered systems. The correlation between the proposed
reduced temperature dependence for $\tau_1$ and the viscosity
models for the amorphous solids supports the conclusion about the
kinetic character of the initiation of the structural ordering in
glasses, where the inherent glassy microscopic dynamics is
predominating over thermodynamic aspects.

Results of this study extend the idea of a unified description of
the nucleation kinetics using scaling relations, which was
originally applied to the analysis of the droplet nucleation rate
data for the vapor-to-liquid transition (see~\cite{Hale_1} and
references to \cite{Hale_2}). The later treatment indicates that
the nucleation rate can be well described by the scaling function
$ \ln (p/p_{coex})/[T_c/T -1 ]^{3/2}$. In this study, we pursue a
similar approach applied to crystallization and define such a
variable, which might provide consistency in comparison of the
\textit{crystal nucleation time} data for different systems. Our
realization differs from the scalings of
Ref.~\cite{Diemand_2,Hale_1}; it is based on the reduced
temperature scale with the fixed control points: the temperature
$\widetilde{T}=0$, the glass transition temperature
$\widetilde{T}_g=0.5$ and the melting temperature
$\widetilde{T}_m=1$ for a considered system. Using this approach
we find a correspondence of the scaled nucleation times as
extracted from simulation and experimental data for the various
systems to a unified power-law dependence. Finally, we note that
because of experimental difficulties in extraction of the
quantitative information about the initial stages of the
crystallization kinetics,  few of the experimental studies cover
the range of supercooling
$(T_{m}-T)/T_{m}>0.6$~\cite{Fokin/Zanotto_2003}.  In this regard,
it could be desirable to verify the suggested scaling law with
additional experimental studies, especially, for the glassy
systems at deep supercooling.

\begin{acknowledgments}
We thank J.-L.~Barrat, D.~Kashchiev, V.N.~Ryzhov, V.V.~Brazhkin,
V.M.~Fokin for helpful discussions. This work was partially
supported by Russian Scientific Foundation (grant RNF
14-13-00676).
\end{acknowledgments}

\bibliographystyle{unsrt}

\begin{thebibliography}{57}

\bibitem{Frenkel_1946} J.~Frenkel, \textit{Kinetic Theory of
Liquids} (Oxford University Press, London, 1946).

\bibitem{Turnbull_1949}  D.~Turnbull, in: J.A.~Prins (Ed.),
\textit{Physics of Non-Crystalline Solids} (North Holland Publishing
Company, Amsterdam, 1965).

\bibitem{Skripov_1974} V.P.~Skripov, \textit{Metastable Liquids}
(Wiley, New York, 1974).

\bibitem{Kelton_1991} K.F.~Kelton, Solid State Phys. {\bf 45}, 75
(1991).

\bibitem{Debenedetti_1996} P.G.~Debenedetti, \textit{Metastable
Liquids. Concepts and Principles} (Princeton Univ. Press, Princeton,
1996).

\bibitem{Kashchiev_Nucleation_2000} D.~Kashchiev,
\textit{Nucleation: Basic Theory with Appplications}
(Butterworth-Heinemann, Oxford, 2000).

\bibitem{Kalikmanov_review} V.I.~Kalikmanov, \textit{Nucleation
Theory, Lecture Notes in Physics vol. 860}(Springer, New York,
2012).

\bibitem{Saika_Voivod_PRL_2011} I.~Saika-Voivod, R.K.~Bowles, and
P.H.~Poole, Phys. Rev. Lett. {\bf 103}, 225701 (2009).

\bibitem{Mokshin/Barrat_PRE_2010} A.V.~Mokshin, J.-L.~Barrat,
Phys. Rev. E {\bf 82}, 021505 (2010).

\bibitem{Mokshin/Barrat_PRE_2008} A.V.~Mokshin, J.-L.~Barrat,
Phys. Rev. E {\bf 77}, 021505 (2008).

\bibitem{Kerrache_PRB_2011_1} A.~Kerrache, N.~Mousseau and
L.J.~Lewis, Phys. Rev. B {\bf 83}, 134122 (2011).

\bibitem{Kerrache_PRB_2011_2} A.~Kerrache, N.~Mousseau and
L.J.~Lewis, Phys. Rev. B {\bf 84}, 014110 (2011).

\bibitem{Heyes_JCP_2012} D.M.~Heyes, E.R.~Smith, D.~Dini,
H.A.~Spikes and T.A.~Zaki, J. Chem. Phys. {\bf 136}, 134705 (2012).

\bibitem{Mokshin/Galimzyanov/Barrat_PRE_2013} A.V.~Mokshin,
B.N.~Galimzyanov and J.-L.~Barrat, Phys. Rev. E {\bf 87}, 062307
(2013).

\bibitem{Durschang} B.R.~Durschang, G.~Carl, C.~R\"{u}ssel and
I.~Gutzow, Berichte der Bunsengesellschaft f\"{u}r physikalische
Chemie {\bf 100}, 1456 (1996).

\bibitem{Gutzow} I.~Gutzow, C.~R\"{u}ssel, and B.~Durschang,
J. Mater. Sci. {\bf 32}, 5405 (1997).

\bibitem{Xing_JAP_2002} P.F.~Xing, Y.X.~Zhuang, W.H.~Wang,
L.~Gerward and J.Z.~Jiang, J. Appl. Phys. {\bf 91}, 4956 (2002).

\bibitem{Yang_JPCM_2008} C.~Yang \textit{et. al} J. Phys.:
Condens. Matter {\bf 20}, 015201 (2008).

\bibitem{Niss_JCP_2008} K.~Niss \textit{et. al}, J. Chem. Phys.
{\bf 129}, 194513 (2008).

\bibitem{Tanguy_2012} B. Mantisi, A. Tanguy, G. Kermouche, E.
Barthel, Eur. Phys. J. B {\bf 85}, 304 (2012).

\bibitem{Zanotto_2002} E.D.~Zanotto, V.M.~Fokin, Phil. Trans. R
Soc. Lond. A {\bf 361}, 591 (2002).

\bibitem{Kalinina_1986} A.M.~Kalinina, V.N.~Filipovich,
V.M.~Fokin, G.A.~Sycheva, in: Proc. XIV Int. Cong. on Glass, New
Delhi {\bf 1}, 366 (1986).

\bibitem{Muller_2000} R.~M\"uller, E.D.~Zanotto, V.M.~Fokin, J.
Non-Cryst. Solids {\bf 274}, 208 (2000).

\bibitem{Fokin/Zanotto_2003} V.M.~Fokin, E.D.~Zanotto,
J.W.P.~Schmelzer, J. Non-Cryst. Solids {\bf 321}, 52 (2003).

\bibitem{Fokin/Yuritsyn/Zanotto_review} V.M.~Fokin, E.D.~Zanotto,
N.S.~Yuritsyn, J.W.P.~Schmelzer, J. Non-Cryst. Solids {\bf 352},
2681 (2006).

\bibitem{James_1989} P.F.~James, in: M.H.~Lewis (Ed.),
\textit{Glasses and Glass-Ceramics} (Chapman and Hall, London,
1989).

\bibitem{Zanotto_1987} E.D.~Zanotto, J. Non-Cryst. Solids {\bf
89}, 361 (1987).

\bibitem{Deubener_2000} J.~Deubener, J. Non-Cryst. Solids {\bf
274}, 195 (2000).

\bibitem{Shevkunov} S. V. Shevkunov, Colloid Journal {\bf 75}, 444
(2013).

\bibitem{Binder_scaling_1976}  K.~Binder and D.~Stauffer, Adv.
Phys. {\bf 25}, 343 (1976).

\bibitem{Hale_1} B.N.~Hale, Phys. Rev. A {\bf 33}, 4156 (1986).

\bibitem{Hale_1_2} B.N.~Hale, J. Chem. Phys. {\bf 122}, 204509 (2005).

\bibitem{Hale_2} B.N.~Hale and M.~Thomason, Phys. Rev. Lett. {\bf
105}, 046101 (2010).

\bibitem{Diemand_1} J.~Diemand, R.~Ang\'{e}lil, K.K. Tanaka, and H.
Tanaka, J. Chem. Phys. {\bf 139}, 074309 (2013).

\bibitem{Diemand_2} K.K. Tanaka, J. Diemand, R. Ang\'{e}lil, and H.
Tanak, J. Chem. Phys. {\bf 140}, 194310 (2014).

\bibitem{Hale_3}
B.N.~Hale, Lecture Notes in Physics {\bf 309}, 323 (1988).

\bibitem{Angell_Leningrad_1989} C.A.~Angell, C.A.~Scamehorn,
D.J.~List, and J.~Kieffer, \textit{Proceedings of XV International
Congress on Glass} (Leningrad, 1989).

\bibitem{Angell_1995}  C.A. Angell, Science {\bf 267}, 1924 (1995).

\bibitem{Dzugutov_1992} M.~Dzugutov, Phys. Rev. A {\bf 46}, R2984
(1992).

\bibitem{Dzugutov_2002} M.~Dzugutov, S.I.~Simdyankin,
F.H.M.~Zetterling, Phys. Rev. Lett. {\bf 89}, 195701 (2002).

\bibitem{Rowlinson_1969} J.S.~Rowlinson, \textit{Liquid and Liquid
Mixtures} (Butterworths, London, 1969).

\bibitem{Toxvaerd_Pedersen_2009} S.~Toxvaerd, U.R.~Pedersen,
T.B.~Schroder, and J.C.~Dyre, J. Chem. Phys. {\bf 130}, 224501
(2009).

\bibitem{Roth_PRB_2005} J.~Roth, Phys. Rev. B {\bf 72}, 014125
(2005).

\bibitem{Hansen_McDonald_2006} J.P.~Hansen, I.R.~McDonald,
\textit{Theory of Simple Liquids} (Academic Press, London, 2006).

\bibitem{Kob_Andersen_1993} W.~Kob and H.C.~Andersen, Phys. Rev.
E {\bf 51}, 4626 (1993); \textit{ibid}. {\bf 52}, 4134 (1995).

\bibitem{Ryzhov}  R.E. Ryltsev, N. M. Chtchelkatchev, V. N. Ryzhov,
Phys. Rev. Lett. {\bf 110}, 025701 (2013).

\bibitem{Auer_Frenkel_2004} S.~Auer and D.~Frenkel, J. Chem.
Phys. {\bf 120}, 3015 (2004).

\bibitem{Mokshin_Barrat_2009} A.V.~Mokshin and J.-L.~Barrat, J.
Chem. Phys. {\bf 130}, 034502 (2009).

\bibitem{Wolde-Frenkel}  P.R. ten Wolde, M.J. Ruiz-Montero, and D. Frenkel, J.
Chem. Phys. {\bf 104}, 9932 (1996).

\bibitem{Steinhardt_Nelson_1983} P.J.~Steinhardt, D.R.~Nelson,
and M.~Ronchetti, Phys. Rev. B {\bf 28}(2), 784 (1983).

\bibitem{Mokshin/Galimzyanov_JCP_2014} A.V.~Mokshin,
B.N.~Galimzyanov, J. Chem. Phys. {\bf 140}, 024104 (2014).

\bibitem{Mokshin_Galimzyanov_JPCB_2012} A.V.~Mokshin,
B.N.~Galimzyanov, J. Phys. Chem. B {\bf 116}, 11959 (2012).

\bibitem{Hanggi_RMP_1990}  P. H\"anggi, P. Talkner, and M. Borkovec, Rev. Mod. Phys.
{\bf 62}, 251 (1990).

\bibitem{Trudu/Parinello} F.~Trudu, D.~Donadio, and
M.~Parrinello, Phys. Rev. Lett. {\bf 97}, 105701 (2006).

\bibitem{Saika_Voivod_JCP_2007} E.~Mendez-Villuendas,
I.~Saika-Voivod, R.K.~Bowles, J. Chem. Phys. {\bf 127}, 154703
(2007).

\bibitem{Bartell_JCP_2007} L.S.~Bartell and D.T.~Wu, J. Chem.
Phys. {\bf 127}, 174507 (2007).

\bibitem{Khusnutdinoff/Mokshin_PhA_2012} R.M. Khusnutdinoff, A.V.
Mokshin, Physica A {\bf 391}, 2842 (2012).

\bibitem{Khusnutdinoff/Mokshin_JNCS_2011} R.M. Khusnutdinoff, A.V.
Mokshin, J. Non-Cryst. Solids {\bf 357}, 1677 (2011).

\bibitem{Fokin/Zanotto_JNCS_2000} V.M.~Fokin, E.D.~Zanotto, J.
Non-Cryst. Solids {\bf 265}, 105 (2000).

\bibitem{Grujic_2009} S.R.~Gruji\'{c}, N.S.~Blagojevi\'{c},
M.B.~To\v{s}i\'{c}, V.D.~\v{Z}ivanovi\'{c}, J.D.~Nikoli\'{c},
Ceramics -- Silik\'{a}ty {\bf 53}, 128 (2009).

\bibitem{Shneidman} V.A.~Shneidman, E.V.~Goldstein, J. Non-Cryst.
Solids {\bf 351}, 1512 (2005).

\bibitem{Kashchiev_1969} D.~Kashchiev, Surf. Sci. {\bf 14}, 209
(1969).

\bibitem{Seeton} C.J.~Seeton, Tribology Letters {\bf 22}, 67
(2006).

\bibitem{Allan_PNAS} J.C.~Mauro, Y.~Yue, A.J.~Ellison,
P.K.~Gupta, D.C.~Allan, PNAS {\bf 106}, 19780 (2009).

\bibitem{Gotze_book_2009} W.~G\"{o}tze,  \textit{Complex Dynamics
of Glass-Forming liquids} (Oxford: Oxford University Press, 2009).

\bibitem{Bottinga} Y.~Bottinga, P.~Richet, A.~Sipp, American
Mineralogist {\bf 80}, 305 (1995).

\bibitem{Avramov/Milchev_1988} I.~Avramov, A.~Milchev, J.
Non-Cryst. Solids {\bf 104}, 253 (1988).

\bibitem{van_Megen_Nature_1993} W.~van Megen and S.M.~Underwood,
Nature {\bf 362}, 616 (1993).

\bibitem{Cavagna_2003} A.~Cavagna, I.~Giardina and T.S.~Grigera,
EPL {\bf 61}, 74 (2003).

\bibitem{Cavagna_2007} A.~Cavagna, A.~Attanasi and J.~Lorenzana,
Phys. Rev. Lett. {\bf 95}, 115702 (2005).

\bibitem{Cahn/Hilliard_JCP_1959} J.W.~Cahn and J.E.~Hilliard, J.
Chem. Phys. {\bf 31}, 688 (1959).

\bibitem{Pan_2006}  A.C.~Pan, T.J.~Rappl, D.~Chandler and
N.P.~Balsara, J. Phys. Chem. B {\bf 110}, 3692 (2006).

\bibitem{Bagchi_2007} P.~Bhimalapuram, S.~Chakrabarty and
B.~Bagchi, Phys. Rev. Lett. {\bf 98}, 206104 (2007).

\bibitem{Sanz_PRL_2011} E.~Sanz, C.~Valeriani, E.~Zaccarelli,
W.C.K.~Poon, P.N.~Pusey and M.E.~Cates, Phys. Rev. Lett. {\bf 106},
215701 (2011).

\end{thebibliography}

\vfill\eject

\end{document}